\documentclass[11pt]{article}
\usepackage{amsmath,amssymb,tabularx,setspace,color,graphicx,multirow,booktabs,tabularx,setspace,color,graphicx}
\usepackage{multirow}
\usepackage[margin=3.5cm]{geometry}
\usepackage{hyperref}

\makeatletter
\renewcommand\section{\@startsection {section}{1}{\z@}%
                                   {-3.5ex \@plus -1ex \@minus -.2ex}%
                                   {2.3ex \@plus.2ex}%
                                   {\normalfont\large\bfseries}}
\makeatother

\usepackage{setspace}
\usepackage{amsmath}
\usepackage{amsthm}
\usepackage{amssymb}
\usepackage{graphicx}
\usepackage{natbib}
\usepackage{amsmath}
\usepackage{amsthm}
\usepackage{amssymb}
\usepackage{graphicx}
\usepackage{natbib}
\begin{document}
\doublespace
\begin{center}
	{\Large {\bf Comparison of cause specific rate functions of panel count data with multiple modes of recurrence}}
\end{center}

\vspace{.1in}
\begin{center}
    Sankaran P. G.$^{a}$, Ashlin Mathew, P. M.$^{b}$ and Sreedevi E. P.$^{c}$ \footnote{ Corresponding Author Email: sreedeviep@gmail.com } \\
	$^{a}$ Cochin University of Science and Technology, Cochin.\\
	$^{b}$ St. Thomas College (Autonomous), Thrissur.\\
	$^{c}$ SNGS College, Pattambi. .
\end{center}

\begin{abstract}
	
	Panel count data refer to the data arising from studies concerning recurrent events where study subjects are observed only at distinct time points. If these study subjects are exposed to recurrent events of several types, we obtain panel count data with multiple modes of recurrence. In the present paper, we propose a nonparametric test for comparing cause specific rate functions of panel count data with more than one mode of recurrence.  The test can also be employed to assess whether the competing modes of recurrence are affecting the recurrence times identically. We carry out simulation studies  to evaluate the performance of the test statistic in a finite sample setup. The proposed test is illustrated using two real life  panel count data sets, one arising from a medical follow up study on skin cancer chemo prevention trial and the other on a warranty database for a fleet of automobiles.
\end{abstract}

\section{Introduction}
\par
Lifetime data analysis often includes studies concerning the recurrent rates or patterns of some events which can occur repeatedly. The recurrent events can be further classified in terms of observation scheme. When each subject is monitored continuously, it provide the exact occurrence times of all events. Such data are usually referred as recurrent event data \citep{cook2007statistical}. But when the study subjects are examined only at discrete time points, the number of occurrence of the events between consecutive observation times are only available; the exact recurrence times remain unknown. This type of data is termed as panel count data  (\citealp{kalbfleisch1985analysis}, \citealp{sun2009analyzing}, \citealp{zhao2011nonparametric}). The panel count data could occur for various reasons.  For example, in many situations continuous observation is too expensive or impossible or it may not be practical to conduct continuous follow-ups of the subjects under study. It can be noted that, the number of observation time points and observation times may vary for each subject. Panel count data is also termed as interval count data or interval censored recurrent event data (\citealp{lawless1998analysis}, \citealp{thall1988analysis}). If each subject is observed only once, the number of recurrences of the event up to the observation time is only available. This special case of panel count data is commonly known as current status data.
\par The two important frame works for the analysis of panel count data focuses on the rate function and mean function of the underlying recurrent event process. \cite{thall1988analysis}  and \cite{lawless1998analysis}  considered the analysis of panel count data using rate functions. An estimator for the mean function based on isotonic regression theory was developed by \cite{sun1995estimation}. \cite{wellner2000two} discussed likelihood  based nonparametric estimation methods for the mean function and proposed a nonparametric maximum likelihood estimator (NPMLE) and a nonparametric  maximum pseudo likelihood estimator (NPMPLE) for the same. They also showed that NPMPLE is exactly the one studied in \cite{sun1995estimation}. Some recent research works in this area include \cite{zhou2017joint}, \cite{xu2018joint}, \cite{wang2019quantile}, \cite{jiang2020robust} and \cite{wang2020bayesian} among others.


\par When an  individual (subject) in the study is exposed to the risk of recurrence due to  several types of events at each point of observation, we obtain panel count data with  multiple modes of recurrence. Such data naturally arise from survival and reliability studies where the interest is focused on the recurrence of competing events which can be observed only at discrete time points. For example, consider the data on skin cancer chemoprevention trial discussed in Sun and Zhao (2013). The cancer recurrences of 290 patients with a history of non-melanoma skin cancers are observed at different monitoring times. The types of cancers are classified into basal cell carcinoma and squamous cell carcinoma and the recurrences due to both types of cancers at each monitoring time are observed for each individual. Covariate information on age, gender, number of prior tumours and DFMO status is also observed for each individual. Accordingly, we have panel count data with multiple modes of recurrence. A detailed analysis of the data is given in Section 4.

Even though recurrent event data exposed to multiple modes of recurrence is studied by many authors in literature \citep{cook2007statistical}, panel count data with multiple modes of recurrence is less explored in literature.  \cite{sreedevi2020nonparametric}  derived an expression for the cause specific mean functions and developed a nonparametric test for comparing the effect of different causes on recurrence times based on the developed estimators. \cite{sankaran2020cause} considered non parametric estimation of cause specific rate functions and studied their properties. When study subjects are exposed to multiple modes of failure/recurrence, it is important to test whether the effect of different causes/modes are identical on the lifetime \citep{gray1988class}.  Many authors including \cite{aly1994some} and \cite{sankaran2010quantile} addressed the above testing problem for right censored data. When the current status data is only available, \cite{sreedevi2012nonparametric} developed a test for independence of time to failure and cause of failure. Comparison of cumulative incidence functions of current status data with continuous and discrete observation times is studied by \cite{sreedevi2014nonparametric} and \cite{sreedevi2019comparison} respectively. Even though current status data can be considered as a special case of panel count data, the estimation procedures are different for both data types and the aforementioned tests cannot be used in the present situation.

\par The test proposed by \cite{sreedevi2020nonparametric} can be used for comparing the mean functions of panel count data with more than one recurrence mode. But there are several advantages in using rate functions for the analysis of panel count data compared to mean functions. Mainly, fewer assumptions are  only required for models based on rate functions. In addition, rate functions are not constrained with the non decreasing property of mean functions and hence it is easy to understand the changing recurrence patterns with rate functions. This motivated us to propose a test to compare the cause specific rate functions proposed by \cite{sankaran2020cause}. Our test is also potent to compare the effect of different recurrence modes on recurrence time for panel count data.

\par The paper is organized as follows. In the Section 2, we discuss the estimation of the cause specific rate functions and then propose a non parametric test to compare the rate functions of panel count data with multiple modes of recurrence. We also discuss the asymptotic properties of the proposed test statistic. In Section 3, we report the results of the simulation study conducted to evaluate the performance of proposed test in finite samples. We illustrate the practical usefulness of the method by applying it to two real data sets in Section 4. Finally, Section 5 summarizes the major conclusions of the study with a discussion of future works.

\vspace{.1in}

\section{Inference procedures}
We study cause specific rate functions and their properties in detail in this section. Further a non parametric test for comparing cause specific rate functions is presented.
\subsection{Cause specific rate functions}
Consider a  study on $n$ individuals from a homogeneous population which are exposed to the recurrent events due to $\{1,2,...,J\}$  possible causes. Assume that the event process is observed only at a sequence of random monitoring times. Consequently, the counts of the event recurrences due to each cause  in between the observation times are only available; the exact recurrence times remain unknown. As a result, we observe the  cumulative number of recurrences upto every observation time due to each cause. Define a counting process $ N_j=\{N_j(t); t\ge 0\}$ where $N_j(t)$ denote the number of recurrences of the event due to cause $j$ upto time $t$.
Define $\mu_j(t)=E(N_j(t))$ as the mean function of the recurrent event process due to cause $j$ which are termed as cause specific mean functions. Define $r_j(t)dt=d\mu_j(t)=EdN_j(t)$ as the rate function of the recurrent event process due to cause $j$, for  $j=1,2,...,J$.  $r_j(t)$ is referred to as the cause specific rate function. By studying cause specific rate functions, one can easily understand the difference in recurrence patterns due to various causes (modes) of recurrence.

\par Note that the number of observation times as well as observation time points may be different for each individual.  Let $M_{i}$ be an integer valued random variable denoting the number of observation times for $i=1,2,..,n$. Also let $T_{i,p}$ denote the $p^{th}$ observation time for $i^{th}$ individual for $p=1,2,..M_i$ and $i=1,2,..,n$. Assume that the number of recurrences due to different causes are independent of number of observation times as well as observation time points.  Let $N_{i,p}^{j}$ denote the number of recurrences of the event observed  for $i^{th}$ individual due to cause $j$ , for $p=1,2,...,M_{i}$, $i=1,2,...,n$ and $j=1,2,...,J$. Now  we observe $n$ independent and identically distributed copies of  $\{M_{i},T_{i,p},N_{i,p}^{1},...,N_{i,p}^{J}\}$, $p=1,2,...,M_{i}$.
The observed data will be of the form $\{m_{i},t_{i,p},n_{i,p}^{1},...,n_{i,p}^{J}\}$, $p=1,2,...,m_{i}$ and $i=1,2,...,n$.

\par \cite{sankaran2020cause} introduced various estimators for cause specific rate functions and established their practical utility through numerical illustrations. The empirical estimators for the cause specific rate functions $r_j(t)$'s are defined as
\begin{equation}
\widehat{r_j(t)}=\frac{\sum_{i=1}^{n}\left[\sum_{p=1}^{m_i}\frac{
		(n_{i,p}^{j}  - n_{i,p-1}^{j})   I(t_{i,p}  < t \leq t_{i,p-1} )}{(t_{i,p} - t_{i,p-1}  )}\right]}{\sum_{i=1}^{n}(t\le t_{i,p} )}~~~~j=1,2,...,J.
\label{eq111}
\end{equation}
In this definition, the numerator gives the average  number of recurrences for subject $i$  due to cause $j$ and denominator is the number of individuals at risk at time $t$. Hence the estimators $\widehat {r_j(t)}$'s are the average of rate functions due to cause $j$ over all individuals.  The cause specific mean functions can be directly estimated from Eqn \eqref{eq111}. When $J=1$, Eqn \eqref{eq111} reduces to the empirical estimator of the rate function given in Sun and Zhao (2013) and the expression is given by
\begin{equation}
\widehat{r(t)}=\frac{\sum_{i=1}^{n}\left[\sum_{p=1}^{m_i}\frac{
		(n_{i,p}  - n_{i,p-1})   I(t_{i,p}  < t \leq t_{i,p-1} )}{(t_{i,p} - t_{i,p-1}  )}\right]}{\sum_{i=1}^{n}(t\le t_{i,p} )}
\label{overall}
\end{equation}
where $n_{i,p}$ is the denote the number of recurrences of the event observed  for $i^{th}$ individual due to all possible modes of recurrence up to time $p$, for $p=1,2,...,M_{i}$, $i=1,2,...,n$.
By definition, $\widehat{r(t)}=\sum_{j=1}^{J}\widehat{r_j(t)}$. In practice, the estimators of cause specific rate functions presented in Eqn (\ref{eq111}) changes only at the observed time points. Accordingly, \cite{sankaran2020cause} proposed a smoothed version of the estimators of cause specific rate functions using kernel estimation techniques and also studied the asymptotic properties.

Let $K(t)$ be a non-negative kernel function symmetric about $t=0$ with $\int_{-\infty}^{\infty}K(t)dt=1$. Also let $h_n>0$ be the bandwidth parameter. Let $b_{1}<b_{2}<...<b_{l}$ are the distinct observed time points in the set $\{T_{i,p}, ~p=1,2,...,M_{i},~ i =1,2,...,n$\}. Define $\widehat{r_{qj}}=\widehat{r_j(b_q)}$, for $q=1,2,...,l$, $j=1,2,...,J$. Now, the kernel estimators of $r_j(t)$'s are given as
\begin{equation}
\widehat{r^{*}_{j}(t)}= \sum_{q=1}^{l} w_q(t)  \widehat{r_{qj}}~~~~j=1,2,...,J.
\label{222}
\end{equation}
where
\begin{equation*}
w_q(t)=\frac{w^*_q(t,h_n)}{\sum_{u=1}^{l}w^*_u(t,h_n)} ~~~~  q=1, 2,...,l.
\end{equation*}
and
\begin{equation*}
w^*_q(t,h_n)=h_n^{-1}K\left(\frac{t-b_q}{h_n}\right)
\end{equation*}
with
\begin{equation*}
K(t)=(2\pi)^{-1/2}\text{exp}(-t^{2}/2).
\end{equation*}
The smoothed estimators $\widehat{r^{*}_{j}(t)}$ of the cause specific rate functions are weighted average of $\widehat{r_{j}(t)}$'s.
Smoothed estimators of over all rate functions can also be constructed in similar way ( Sun and Zhao, 2013). Clearly, $\widehat{r^{*}(t)}=\sum_{j=1}^{J}\widehat {r^{*}_j(t)}$, where $\widehat{r^{*}(t)}$ is the kernel estimator of the overall rate function. In practice, the bandwidth $h_n$ for which the MSE is minimum is selected to employ smoothing.

\par The asymptotic properties of the estimators $\widehat{r^{*}_{j}(t)}$'s are studied and derived in \cite{sankaran2020cause}. Without loss of generality, assume that the kernel function $K(x)$ satisfies the following mild regularity conditions.\\
C1 : $K(x)$ is bounded ie sup\{$K(x), x \in R \}<\infty $\\
C2 : $|xK(x)| \to 0$ as $|x| \to \infty$  \\
C3 : $K(x)$ is symmetric about 0, ie $K(-x)=K(x)$, $x \in R$\\
Also suppose that, as $n \to \infty$ the bandwidth parameter $h_n$ satisfies the conditions
(i) $h_n \to 0$ (ii) $nh_n \to \infty$ and (iii) $nh^{2}_n \to \infty$. \\
Under the assumptions C1, C2 and C3, \cite{sankaran2020cause}  showed that for fixed $t$, the estimators  $\widehat{r^{*}_{j}(t)}$'s  are asymptotically normal with mean $\lambda_j(t)=E(\widehat{r^{*}_{j}(t)})$ and standard deviation $\sigma_j(t)=\text{s.d}(\widehat{r^{*}_{j}(t)})$ for $j=1,2,...,J$.

\subsection{Test statistic}

In this study, we focus on comparing the cause specific rate functions due to various recurrence modes. This may be helpful in selecting the appropriate treatment for a group of patients in a clinical study or to evaluate a newly introduced system in reliability experiments. To develop a test statistic, we now consider the hypothesis,
$$H_0:  {r_{j}(t)}= {r_{j'}(t)}~\text{for all } t>0,~~j\ne j'=1,2,...,J$$
 against
 \begin{equation}
  H_1:  {r_{j}(t)} \ne {r_{j'}(t)}~\text{for some } t>0~~ \text{and }~~j\ne j'=1,2,...,J.
  \label{333}
 \end{equation}
Since $r(t)=\sum_{j=1}^{J}r_j(t)$, the above hypothesis can also be written as
$$H_0:  {r_{j}(t)}= \frac{r(t)}{J}~\text{for all } t>0,~~j\ne j'=1,2,...,J$$
 against
\begin{equation}
H_1:  {r_{j}(t)} \ne \frac{r(t)}{J}~\text{for some } t>0~~ \text{and }~~j\ne j'=1,2,...,J.
\end{equation}
To test $H_0$ against $H_1$, we choose $\widehat{r^{*}_{j}(t)}$ as the smoothed estimators for the cause specific rate functions defined in Eqn \eqref{222}. A smoothed estimator for the overall rate function $r(t)$ specified in Eqn \eqref{overall} is constructed by omitting the information on the mode of recurrence. Let $ \widehat{r^*(t)}$  denote smoothed estimator of overall rate function. A similar procedure of estimating the overall mean function by ignoring the cause of recurrence information is used in \cite{sreedevi2020nonparametric} for comparing cause specific mean functions.

To develop a test statistic for comparing cause specific rate functions, consider the function
\begin{equation}
v_{j}(t)=\int_{0}^{t} w(u)\left[\widehat{r^{*}_{j}(u)}-\frac{\widehat{r^{*}(u)}}{J}\right] \mathrm{d} u \quad \text { for all } j=1,2, \ldots, J
\end{equation}
where $w(.)$ is an appropriate data dependent weight function which is used to compensate the effect of censoring. The weight functions are also employed to increase the efficiency of the test statistic and to set it asymptotically distribution free \citep{pepe1993kaplan}. The function $v_j(.)$ is similar to the one proposed by \cite{sreedevi2020nonparametric} to compare the cause specific mean functions of panel count data. Now to test the null hypothesis given in \eqref{333}, we propose the test statistic
\begin{equation}
Z(\tau)=v^{\prime}(\tau) \hat{\sum}(\tau)^{-1} v(\tau)
\label{eq777}
\end{equation}
where $\tau$ is the largest monitoring time in the study and $v(\tau)=\left(v_{1}(\tau), \ldots, v_{k}(\tau)\right)^{\prime} ; \sum {(\tau)^{-1}}$  is the generalized inverse $\hat\sum(\tau)$, where $\hat\sum(\tau)$ is a consistent estimator of $\sum(\tau)$, the variance-covariance matrix of $v(\tau)$. The matrix $\sum(\tau)$ involves variances of $\widehat{r^*_{j}(t)}$ and $\widehat{r(t)}$ and covariances between $\widehat{r^*_{j}(t)}$ and $\widehat{r^*_{j'}(t)}$ for $j \neq j^{\prime}=1,2, \ldots, J$ and that between $\widehat{r^*_{j}(t)}$ and $\widehat{r(t)}$. Bootstrap procedure is used to find the estimate of the variance-covariance matrix, since the expression for $\sum(\tau)$ is complex.
From the asymptotic properties of the kernal estimators of cause specific rate functions discussed in \cite{sankaran2020cause} it follows that, under $H_{0}$ for any $t>0$, the distribution of $v(\tau)=\left(v_{1}(\tau), \ldots, v_{J}(\tau)\right)^{\prime}$ can be asymptotically approximated by a $J-$ variate normal with mean zero vector and variance-covariance matrix $\sum(\tau)$, where $\tau$ is the largest monitoring time in the study.
Now to find the asymptotic distribution of the test statistic Z($
\tau$) given in Eqn.\eqref{eq777}, consider the quantity

\begin{equation*}
v_{j}(t)=\int_{0}^{t} w(u)\left[\widehat{r^{*}_{j}(u)}-\frac{\widehat{r^{*}(u)}}{J}\right] \mathrm{d} u \quad \text { for all } j=1,2, \ldots, J
\end{equation*}
which can be written as

\begin{equation*}
\begin{aligned}
v_{j}(t)=& \int_{0}^{t} w(u)\left[\widehat{r^{*}_{j}(u)}-r(u)\right] \mathrm{d}(u)+\int_{0}^{t} w(u)\left[r_{j}(u)-\frac{r(u)}{J}\right] \mathrm{d} u \\
&+\int_{0}^{t} w(u)\left[\frac{r(u)}{J}-\frac{\widehat{r^{*}(u)}}{J}\right] \mathrm{d} u, \quad j=1,2, \ldots J
\end{aligned}
\end{equation*}
Now under $H_0$, $ {r_{j}(t)}= {r(t)}/{J}~\text{for all } t$, we get
\begin{equation*}
v_{j}(t)=\int_{0}^{t} w(u)\left[\widehat{r^{*}_{j}(u)}-r(u)\right] \mathrm{d} u+\int_{0}^{t} w(u)\left[\frac{r_{j}(u)}{J}-\frac{\widehat{r^{*}(u)}}{J}\right] \mathrm{d} u, \quad j=1,2, \ldots, J
\end{equation*}
Accordingly, under the regularity conditions stated above, the quadratic form $Z(\tau)$ follows a $\chi^{2}$ distribution with $(J-1)$ degrees of freedom. We reject $H_{0}$, if $Z(t) \geq \chi_{\alpha, (J-1)}^{2}$ where $\chi_{\alpha, (J-1)}^{2}$ is the ordinate value of chi-square distribution with $(J-1)$ degrees of freedom at $\alpha$ level.

\section{Simulation studies}

We conduct simulation studies to evaluate the performance of the proposed test statistic in finite samples. The situation with two modes of recurrence is considered here.
A real life situation in medical follow-up study is taken as a model to generate panel count data of the form
$\{m_{i},t_{i,p},n_{i,p}^{1},n_{i,p}^{2}\}$ for $p=1,2,...,m_{i}$ and $i=1,2,...,n$. The number of observation times $m_{i}$ for each individual is generated from a discrete uniform distribution $U(1,10)$ for $i=1,2,...,n$. Thus the maximum number of observations for each individual is restricted upto 10. Then we generated gap times between each observation from uniform distribution $U(0,5)$. The discrete observation time points $t_{i,p}$ for $p=1,2,...,m_i$ and $i=1,2,...,n$ are generated using the above mentioned time gaps. A bivariate Poisson distribution with parameters $(\theta_1, \theta_2, \theta_3)$ is employed to generate recurrent processes $n_{i,p}^{1}$ and $n_{i,p}^{2}$. The joint mass function of the bivariate Poisson distribution with parameters $(\theta_1, \theta_2, \theta_3)$ is given by
\begin{equation}
f(x,y)=\exp\{-(\theta_1+\theta_2+\theta_3)\} \frac{{\theta_1}^x}{x!}\frac{{\theta_2}^y}{y!}\sum_{k=0}^{min(x,y)}{x \choose k}{y \choose k}k! \left(\frac{\theta_3}{\theta_1 \theta_2}\right)^k.
\end{equation}

\begin{table}
	\centering
	\caption{Empirical Type I error and power of the test in percentage for the weight functions $w(.)=1,w(.)= n$ and $w(.)=\widehat{r^*(t)}$}
	\begin{tabular}{lclclclclclclclclcl}
			\hline
		&   & \multicolumn{3}{l}{ ~~~~~~~~~~n~~~} &         &   & \multicolumn{3}{l!{\color{black}}}{~~~~~~~~~n}  \\
\hline 
		( $\theta_1,\theta_2,\theta_3$)     & $\alpha$ & 100  & 200  & 500      & ($\theta_1,\theta_2,\theta_3$ )     & $\alpha$ & 100  & 200  & 500                            \\\hline 
		& &   &      &              $w(t)=1$          &   &      &      &                                \\  \hline 
		
		(1,1,1) & 5 & 5.8  & 5.4  & 5.1      & (1,1,2) & 5 & 5.6  & 5.2  & 4.9                            \\
		
		& 1 & 2    & 1.7  & 1.3      &         & 1 & 1.7  & 1.4  & 1.1                            \\
		
		(1,2,1) & 5 & 65.8 & 71.4 & 79.5     & (1,2,2) & 5 & 66.8 & 74.8 & 80.7                           \\
		
		& 1 & 63.7 & 67.2 & 73.1     &         & 1 & 65.2 & 73.1 & 75.2                           \\
		
		(1,3,1) & 5 & 74.5 & 81.9 & 86.4     & (1,3,2) & 5 & 81.5 & 87.7 & 92.4                           \\
		
		& 1 & 73.0 & 78.6 & 83.1     &         & 1 & 79.4 & 85.6 & 91.6                           \\
		
		(1,4,1) & 5 & 90.3 & 92.1 & 97.2     & (1,4,2) & 5 & 96.5 & 98.2 & 99.9                           \\
		
		& 1 & 87.4 & 91.8 & 94.5     &         & 1 & 96.8 & 98.2 & 99.1                           \\
		
		(1,5,1) & 5 & 98.9 & 100  & 100      & (1,5,2) & 5 & 100  & 100  & 100                            \\
		
		& 1 & 98.4 & 99.7 & 100      &         & 1 & 99.8 & 100  & 100                            \\
\hline 
	 &   &      &      &      	$w(t)=n$           &   &      &      &                             \\  \hline 
		
		(1,1,1) & 5 & 4.5  & 4.7  & 5.2      & (1,1,2) & 5 & 4.4  & 4.8  & 5.1                            \\
		
		& 1 & 2    & 1.7  & 1.3      &         & 1 & 1.4  & 1.3  & 0.9                            \\
		
		(1,2,1) & 5 & 67.1 & 73.2 & 78.4     & (1,2,2) & 5 & 70.4 & 79.5 & 84.7                           \\
		
		& 1 & 66.7 & 69.2 & 74.1     &         & 1 & 68.1 & 74   & 79                             \\
		
		(1,3,1) & 5 & 79.6 & 83.9 & 86.4     & (1,3,2) & 5 & 85.2 & 89.3 & 94.7                           \\
		
		& 1 & 73.0 & 78.6 & 83.1     &         & 1 & 80.5 & 87.2 & 93.7                           \\
		
		(1,4,1) & 5 & 94.3 & 98.1 & 99.9     & (1,4,2) & 5 & 99.9 & 100  & 100                            \\
		
		& 1 & 87.4 & 96.8 & 97.2     &         & 1 & 99.8 & 99.9 & 100                            \\
		
		(1,5,1) & 5 & 100  & 100  & 100      & (1,5,2) & 5 & 100  & 100  & 100                            \\
		
		& 1 & 100  & 100  & 100      &         & 1 & 99.8 & 100  & 100                            \\ \hline

		 &   &      &      &          $w(t)=\widehat{r^*(t)}$           &   &      &      &                                \\ \hline
		
		(1,1,1) & 5 & 4.7  & 5.2  & 5        & (1,1,2) & 5 & 5.5  & 4.8  & 5.1                            \\
		
		& 1 & 0.7  & 1.2  & 0.9      &         & 1 & 1.3  & 1.2  & 1                              \\
		
		(1,2,1) & 5 & 73.2 & 81.0 & 85.7     & (1,2,2) & 5 & 76.9 & 84.1 & 85.4                           \\
		
		& 1 & 71.1 & 78.9 & 84.3     &         & 1 & 71.0 & 77.2 & 84.2                           \\
		
		(1,3,1) & 5 & 89.5 & 92.5 & 98.4     & (1,3,2) & 5 & 88.8 & 91.4 & 97.5                           \\
		
		& 1 & 83.2 & 88.6 & 96.9     &         & 1 & 85.0 & 87.3 & 96.0                           \\
		
		(1,4,1) & 5 & 99.9 & 100  & 100      & (1,4,2) & 5 & 100  & 100  & 100                            \\
		
		& 1 & 99.7 & 100  & 100      &         & 1 & 99.8 & 99.8 & 100                            \\
		
		(1,5,1) & 5 & 100  & 100  & 100      & (1,5,2) & 5 & 100  & 100  & 100                            \\
		
		& 1 & 100  & 100  & 100      &         & 1 & 100  & 100  & 100                            \\\hline
	\end{tabular}
\end{table}
The marginal distribution of $X$ and $Y$ is Poisson distribution with $E(X)=\theta_{1}+\theta_{3}$, $E(Y)=\theta_{2}+\theta_{3}$ and cov$(X,Y)=\theta_{3}$ gives a measure of dependence between random  variables $X$ and $Y$. \cite{sankaran2020cause} used a similar procedure to generate panel count data with multiple failure modes.

In the above simulation frame work, if we set $\theta_{1}=\theta_{2}$ and assign a non zero value for $\theta_{3}$, it corresponds to  a situation where the  cause specific rate functions are identical. Accordingly, the null hypothesis $H_0$ will be true. When the difference between $\theta_{1}$  and $\theta_2$ increases, the difference between the two rate functions also increase which results in a situation where null hypothesis is false.
Hence the parameter combination with $\theta_1=\theta_2$ gives the type I error of the test and all other choices of parameter combinations give the power of the proposed test. We carry out simulation studies for different combinations of $(\theta_1,\theta_2,\theta_3)$ to calculate the empirical type I error and power of the test. For this purpose, observations of different sample sizes n = 100 or n = 200 or n = 500 are simulated and the process is repeated 1000 times. We employ three different choices of weight functions in this study which are $(i)~w(t)=1$, $(ii)~w(t)=n$, the number of individuals in the study and $(iii )~w(t)=\widehat{r^*(t)}$, the smoothed estimator of overall rate function.

Table 1 gives the type I error and the power of the proposed test statistic in percentage for significance level $\alpha= 0.05$ and $\alpha=0.01$. From Table 1, we can see that type I error of the test approaches the chosen significance level. The test is efficient in terms of power also. Also, as the difference between $\theta_1$ and $\theta_2$ increases,  power of the test also increases.   
\section{Data analysis}
The proposed inference procedures are illustrated using two real life data sets in this section.
\subsection{Skin cancer chemo prevention trial data}
We consider the data arising from the  skin cancer chemo prevention trial given in \cite{sun2013statistical} for demonstration.
The study was conducted to study the effectiveness of the DFMO  (DIfluromethylornithire) drug in reducing new skin cancers in a population with a history of non-melanoma skin cancers, basal cell carcinoma and squamous cell carcinoma. The data consists of 290 patients with a history of non-melanoma skin cancers. The  observation and follow-up times differ for each patient. The data has the counts of two types of recurring events basal cell carcinoma and squamous cell carcinoma which we treat here as two modes of recurrence \citep{sreedevi2020nonparametric}.

In the data set, the number of observations on an individual varies from 1 to 17 and the time of observation varies from 12 to 1766 days. The cause specific rate functions due to basal  cell carcinoma and squamous cell carcinoma are estimated using Eqn (\ref{222}). Further, the proposed procedures are applied to evaluate the test statistic. Table 2 gives the chi square test statistic values of the proposed test statistic for different weight functions. From the value of the test statistics, it is clear that we can reject the null hypothesis and conclude that the rate functions due to basal cell carcinoma  and squamous cell carcinoma are significantly different.

The plots of the kernel estimators with bandwidth parameter value $h_n=1.76 \approx n^{\frac{1}{10}} $ is given in Figure 1. The bandwidth value $h_n=1.76$ is chosen from simulation studies, which minimize the MSE of the estimates.
\begin{figure}[!htb]
	\centering
	\includegraphics[width=0.9\linewidth]{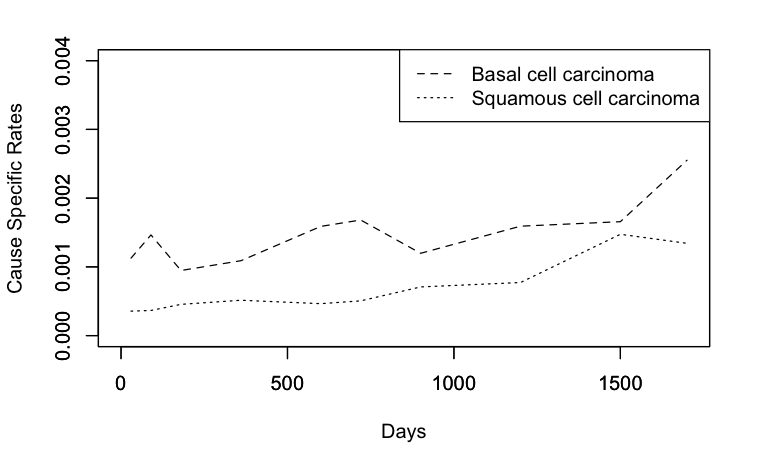}
	\caption{Kernel estimates of cause specific rate functions  due to basal cell carcinoma and squamous cell carcinoma for $h_n=1.76$ }
	\label{fig:skintumor2}
\end{figure}
\begin{table}[ht]
\caption{Test statistic values of the proposed test for different weight functions.}
	\centering
	\begin{tabular}{lccc}
	\hline \\ Weight function & Test statistic & $p$ -value \\
		\hline 1 & $26.97$ & $<.0005$ \\
		$\mathrm{n}$ & $31.92$ & $<.0005$ \\
		$\widehat{r^{*}(.)}$ & $37.74$ & $<.0005$ \\
		\hline
	\end{tabular}
\end{table}

\par From Figure 1, it can be noted that the recurrence rate of basal cell carcinoma is greater than the recurrence rate of squamous cell carcinoma at all time points, which clearly indicates the rejection of $H_0$. Since the rate functions are not monotonic, the change points of recurrence patterns can also be easily identified from the graph.

\subsection{Automobile warranty claims data}
We apply the proposed methods to the automobile warranty claims data studied in  \cite{somboonsavatdee2015parametric}. The data set comprises of recurrent failure history of a fleet of automobiles. The outcome of interest is the repeated mileages at failure for multiple vehicles of a certain model and make, obtained from the warranty claim database and the labour code associated with the failure. In the data, the source and specifics are masked for de-identification purposes. The database
consists of recurrent failure history of 456 vehicles for which a single type I censoring at 3000 miles is considered. Fourteen different labor codes of the warranty claims of each vehicle were recorded with mileage at filing. Due to the absence of a specific description of the component associated with labor code, the grouping was determined on the basis of rate of failures. The fourteen individual labor codes were combined into three broad groups of failure modes FM1, FM2 and FM3, where FM1 comprises of labor codes with shape parameters ranging between 0.2 and 0.36, FM2 covers labor codes with shape parameter estimates between 0.4 and 0.55, whereas FM3 combines the remaining codes that have the slowest rate of growth with shape parameter estimates varying between 0.7 and 0.93. The table IV in \cite{somboonsavatdee2015parametric} presents the data 172 vehicles that have at least one documented record of warranty claim for repair. 
\begin{table}[ht]
\caption{Test statistic values of the proposed test for different weight functions.}
	\centering
	\begin{tabular}{lcc}
	\hline  Weight function & Test statistic & $p$ -value \\ 
		\hline 1 & $49.15$ & $<.0005$ \\
		$\mathrm{n}$ & $68.96$ & $<.0005$ \\
		$\widehat{r^{*}(.)}$ & $79.55$ & $<.0005$ \\ 
		\hline 	
	\end{tabular}
\end{table}
\begin{figure}[!htb]
	\centering
	\includegraphics[width=0.9\linewidth]{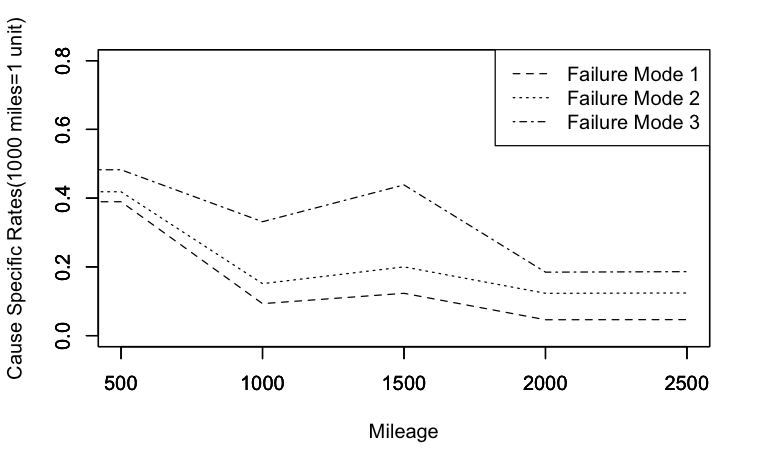}
	\caption{Kernel estimates of cause specific rate functions  due to three modes of failures for $h_n=1.67$ }
	\label{fig:automobile}
\end{figure}
We observed the recurrent failure history data at 1000, 2000 and 3000 mileages at which the number of failures due to each mode are noted, thereby making the recurrent event data as a panel count data with multiple modes of recurrence. The complete data set used in our study is  given in Table 4.

Table 3 gives the chi square test statistic values of the proposed test statistic for different weight functions for automobile warranty data. From the value of the test statistics, it is clear that we can reject the null hypothesis and conclude that the rate functions due to three modes of failure are significantly different.

The plots of the kernel estimators with bandwidth parameter value  $h_n=1.67 \approx n^{\frac{1}{10}} $ is given in Figure 2. The bandwidth value $h_n= 1.67$ is chosen from simulation studies, which minimize the MSE of the estimates. From Figure 2, it can be noted that the recurrence rates of each modes of recurrence (FM1, FM2 and FM3) are distinct at all observed miles, which clearly indicates the rejection of $H_0$.

\section{Conclusion}
In the present paper, we developed non parametric inference procedures for the analysis of panel count data with multiple modes of recurrence based on cause specific rate functions. We proposed a test statistic to test the equality of cause specific rate functions. An extensive simulation study was carried out by generating the data from a bivariate Poisson process to assess the performance of the proposed test in finite samples. Two real life data sets, one from skin cancer chemo prevention trial \citep{sun2013statistical} and other from automobile warranty claims \citep{somboonsavatdee2015parametric} were analysed to demonstrate the practical utility of the procedures.
\par The nature of dependence between time to  failure and cause of failure is important for modelling competing risks data. Even though the problem is studied under right censoring, it is unexplored for panel count data. We can use either cause specific mean functions or cause specific rate functions to tackle this problem.  Works in this direction will be done separately. Regression analysis of panel count data with multiple modes of recurrence  using rate functions is also under investigation.

\begin{center}
	{\bf Acknowledgments}\\
\end{center}
\par The first author would like to thank Science Engineering and Research Board, DST, Government of India and the third author acknowledge the gratitude to Kerala State Council for Science Technology and Environment for the financial support provided to carry out this research work.

\vspace{.2in}


\begin{table}[ht]
 \caption{Automobile Warranty Data}
\begin{tabular}{cccccccccccccccc}
\textbf{ID} & \textbf{MIL} & \textbf{FM1} & \textbf{FM2} & \textbf{FM3} & \textbf{TOTAL} & \textbf{ID} & \textbf{MIL} & \textbf{FM1} & \textbf{FM2} & \textbf{FM3} & \textbf{TOTAL} \\
\textbf{1}  & 1000         & 1            & 1            & 0            & 2              & \textbf{37} & 1000         & 1            & 0            & 0            & 1              \\
\textbf{1}  & 3000         & 1            & 0            & 0            & 1              & \textbf{37} & 2000         & 0            & 0            & 1            & 1              \\
\textbf{2}  & 1000         & 0            & 0            & 2            & 2              & \textbf{38} & 1000         & 1            & 1            & 0            & 2              \\
\textbf{3}  & 3000         & 0            & 0            & 1            & 1              & \textbf{39} & 1000         & 0            & 2            & 0            & 2              \\
\textbf{4}  & 2000         & 0            & 0            & 1            & 1              & \textbf{40} & 1000         & 0            & 2            & 0            & 2              \\
\textbf{5}  & 1000         & 1            & 1            & 1            & 3              & \textbf{41} & 3000         & 0            & 0            & 1            & 1              \\
\textbf{5}  & 2000         & 1            & 0            & 0            & 1              & \textbf{42} & 1000         & 0            & 0            & 1            & 1              \\
\textbf{6}  & 1000         & 0            & 0            & 1            & 1              & \textbf{43} & 1000         & 0            & 0            & 1            & 1              \\
\textbf{7}  & 1000         & 1            & 0            & 0            & 1              & \textbf{44} & 3000         & 0            & 1            & 0            & 1              \\
\textbf{8}  & 1000         & 1            & 0            & 0            & 1              & \textbf{45} & 1000         & 2            & 0            & 0            & 2              \\
\textbf{9}  & 1000         & 0            & 1            & 0            & 1              & \textbf{46} & 1000         & 0            & 1            & 0            & 1              \\
\textbf{10} & 2000         & 0            & 0            & 2            & 2              & \textbf{47} & 1000         & 1            & 0            & 0            & 1              \\
\textbf{11} & 1000         & 1            & 0            & 0            & 1              & \textbf{47} & 3000         & 0            & 1            & 0            & 1              \\
\textbf{12} & 1000         & 1            & 0            & 0            & 1              & \textbf{48} & 1000         & 1            & 1            & 0            & 2              \\
\textbf{13} & 3000         & 0            & 0            & 1            & 1              & \textbf{49} & 1000         & 0            & 1            & 0            & 1              \\
\textbf{14} & 1000         & 0            & 1            & 1            & 2              & \textbf{50} & 1000         & 0            & 0            & 1            & 1              \\
\textbf{15} & 1000         & 0            & 1            & 0            & 1              & \textbf{51} & 3000         & 0            & 0            & 1            & 1              \\
\textbf{15} & 2000         & 0            & 1            & 0            & 1              & \textbf{52} & 1000         & 0            & 0            & 1            & 1              \\
\textbf{16} & 2000         & 0            & 1            & 1            & 2              & \textbf{53} & 2000         & 1            & 0            & 0            & 1              \\
\textbf{16} & 3000         & 0            & 1            & 0            & 1              & \textbf{54} & 1000         & 0            & 1            & 0            & 1              \\
\textbf{17} & 1000         & 1            & 2            & 1            & 4              & \textbf{55} & 1000         & 1            & 0            & 0            & 1              \\
\textbf{17} & 2000         & 1            & 0            & 0            & 1              & \textbf{56} & 1000         & 0            & 1            & 0            & 1              \\
\textbf{18} & 3000         & 0            & 0            & 1            & 1              & \textbf{57} & 1000         & 0            & 2            & 0            & 2              \\
\textbf{19} & 1000         & 0            & 1            & 0            & 1              & \textbf{57} & 2000         & 1            & 0            & 1            & 2              \\
\textbf{20} & 1000         & 1            & 0            & 0            & 1              & \textbf{58} & 1000         & 0            & 0            & 1            & 1              \\
\textbf{21} & 1000         & 0            & 1            & 0            & 1              & \textbf{59} & 1000         & 0            & 1            & 0            & 1              \\
\textbf{22} & 3000         & 0            & 1            & 0            & 1              & \textbf{60} & 1000         & 0            & 1            & 0            & 1              \\
\textbf{23} & 1000         & 1            & 0            & 0            & 1              & \textbf{61} & 2000         & 1            & 0            & 0            & 1              \\
\textbf{24} & 3000         & 1            & 0            & 0            & 1              & \textbf{62} & 1000         & 0            & 1            & 0            & 1              \\
\textbf{25} & 1000         & 0            & 1            & 0            & 1              & \textbf{63} & 2000         & 0            & 0            & 1            & 1              \\
\textbf{26} & 1000         & 1            & 0            & 1            & 2              & \textbf{64} & 1000         & 0            & 0            & 1            & 1              \\
\textbf{26} & 2000         & 1            & 2            & 0            & 3              & \textbf{65} & 1000         & 1            & 0            & 0            & 1              \\
\textbf{26} & 3000         & 0            & 2            & 0            & 2              & \textbf{66} & 1000         & 2            & 1            & 0            & 3              \\
\textbf{27} & 3000         & 0            & 1            & 0            & 1              & \textbf{67} & 1000         & 1            & 0            & 0            & 1              \\
\textbf{28} & 2000         & 0            & 0            & 1            & 1              & \textbf{67} & 3000         & 0            & 0            & 1            & 1              \\
\textbf{29} & 1000         & 1            & 0            & 1            & 2              & \textbf{68} & 1000         & 0            & 1            & 0            & 1              \\
\textbf{30} & 3000         & 0            & 2            & 0            & 2              & \textbf{69} & 2000         & 0            & 1            & 0            & 1              \\
\textbf{31} & 2000         & 0            & 1            & 0            & 1              & \textbf{70} & 1000         & 1            & 0            & 0            & 1              \\
\textbf{32} & 2000         & 0            & 1            & 0            & 1              & \textbf{71} & 1000         & 1            & 0            & 0            & 1              \\
\textbf{33} & 3000         & 0            & 0            & 1            & 1              & \textbf{72} & 2000         & 0            & 0            & 2            & 2              \\
\textbf{34} & 1000         & 0            & 1            & 0            & 1              & \textbf{73} & 1000         & 1            & 0            & 0            & 1              \\
\textbf{35} & 1000         & 0            & 0            & 1            & 1              & \textbf{73} & 2000         & 0            & 0            & 1            & 1              \\
\textbf{35} & 2000         & 1            & 0            & 0            & 1              & \textbf{74} & 1000         & 1            & 0            & 1            & 2              \\
\textbf{36} & 1000         & 0            & 1            & 0            & 1              & \textbf{74} & 2000         & 0            & 0            & 1            & 1
\end{tabular}
\end{table}

\begin{table}[ht]
\begin{tabular}{cccccccccccc}
\textbf{ID}           & \textbf{MIL} & \textbf{FM1} & \textbf{FM2} & \textbf{FM3} & \textbf{TOTAL} & \textbf{ID}           & \textbf{MIL} & \textbf{FM1} & \textbf{FM2} & \textbf{FM3} & \textbf{TOTAL} \\
\textbf{75}  & 1000         & 1            & 0            & 0            & 1              & \textbf{113} & 1000         & 0            & 0            & 1            & 1              \\
\textbf{76}  & 1000         & 0            & 0            & 1            & 1              & \textbf{114} & 1000         & 1            & 0            & 0            & 1              \\
\textbf{77}  & 1000         & 0            & 1            & 1            & 2              & \textbf{115} & 1000         & 0            & 1            & 1            & 2              \\
\textbf{78}  & 1000         & 0            & 1            & 0            & 1              & \textbf{116} & 1000         & 1            & 0            & 0            & 1              \\
\textbf{79}  & 3000         & 0            & 0            & 1            & 1              & \textbf{117} & 2000         & 0            & 1            & 1            & 2              \\
\textbf{80}  & 1000         & 1            & 0            & 0            & 1              & \textbf{118} & 2000         & 0            & 0            & 1            & 1              \\
\textbf{81}  & 1000         & 0            & 0            & 1            & 1              & \textbf{119} & 2000         & 1            & 0            & 0            & 1              \\
\textbf{82}  & 1000         & 1            & 0            & 0            & 1              & \textbf{120} & 1000         & 1            & 0            & 1            & 2              \\
\textbf{83}  & 1000         & 0            & 0            & 1            & 1              & \textbf{121} & 1000         & 0            & 0            & 1            & 1              \\
\textbf{84}  & 2000         & 0            & 0            & 1            & 1              & \textbf{121} & 3000         & 0            & 0            & 1            & 1              \\
\textbf{85}  & 1000         & 0            & 2            & 0            & 2              & \textbf{122} & 1000         & 1            & 0            & 1            & 2              \\
\textbf{86}  & 1000         & 0            & 0            & 1            & 1              & \textbf{123} & 2000         & 0            & 1            & 0            & 1              \\
\textbf{86}  & 2000         & 0            & 2            & 0            & 2              & \textbf{124} & 1000         & 1            & 0            & 0            & 1              \\
\textbf{87}  & 1000         & 1            & 0            & 0            & 1              & \textbf{125} & 2000         & 0            & 0            & 1            & 1              \\
\textbf{88}  & 2000         & 0            & 0            & 1            & 1              & \textbf{126} & 1000         & 2            & 0            & 1            & 3              \\
\textbf{88}  & 3000         & 0            & 0            & 1            & 1              & \textbf{126} & 3000         & 0            & 0            & 2            & 2              \\
\textbf{89}  & 3000         & 1            & 0            & 0            & 1              & \textbf{127} & 2000         & 0            & 0            & 1            & 1              \\
\textbf{90}  & 1000         & 0            & 0            & 2            & 2              & \textbf{128} & 2000         & 0            & 1            & 0            & 1              \\
\textbf{90}  & 3000         & 0            & 0            & 1            & 1              & \textbf{129} & 1000         & 2            & 3            & 1            & 6              \\
\textbf{91}  & 1000         & 0            & 1            & 0            & 1              & \textbf{129} & 2000         & 0            & 0            & 1            & 1              \\
\textbf{92}  & 1000         & 0            & 1            & 0            & 1              & \textbf{130} & 1000         & 0            & 1            & 0            & 1              \\
\textbf{93}  & 1000         & 0            & 0            & 1            & 1              & \textbf{131} & 1000         & 1            & 0            & 0            & 1              \\
\textbf{94}  & 1000         & 1            & 1            & 0            & 2              & \textbf{132} & 3000         & 0            & 0            & 1            & 1              \\
\textbf{95}  & 1000         & 1            & 0            & 0            & 1              & \textbf{133} & 2000         & 1            & 0            & 1            & 2              \\
\textbf{96}  & 2000         & 0            & 0            & 1            & 1              & \textbf{134} & 2000         & 0            & 1            & 1            & 2              \\
\textbf{97}  & 2000         & 0            & 0            & 1            & 1              & \textbf{135} & 1000         & 0            & 0            & 1            & 1              \\
\textbf{98}  & 1000         & 0            & 1            & 0            & 1              & \textbf{136} & 1000         & 0            & 0            & 1            & 1              \\
\textbf{98}  & 2000         & 1            & 1            & 1            & 3              & \textbf{137} & 1000         & 0            & 0            & 1            & 1              \\
\textbf{99}  & 1000         & 1            & 0            & 0            & 1              & \textbf{138} & 1000         & 0            & 0            & 1            & 1              \\
\textbf{100} & 1000         & 1            & 0            & 1            & 2              & \textbf{138} & 3000         & 1            & 0            & 0            & 1              \\
\textbf{101} & 1000         & 0            & 0            & 1            & 1              & \textbf{139} & 1000         & 1            & 0            & 0            & 1              \\
\textbf{102} & 1000         & 1            & 0            & 0            & 1              & \textbf{140} & 1000         & 1            & 0            & 0            & 1              \\
\textbf{103} & 1000         & 1            & 0            & 0            & 1              & \textbf{141} & 3000         & 0            & 1            & 0            & 1              \\
\textbf{104} & 2000         & 0            & 0            & 1            & 1              & \textbf{142} & 1000         & 0            & 1            & 1            & 2              \\
\textbf{105} & 1000         & 1            & 0            & 0            & 1              & \textbf{143} & 1000         & 1            & 0            & 0            & 1              \\
\textbf{106} & 1000         & 0            & 0            & 2            & 2              & \textbf{143} & 3000         & 0            & 0            & 1            & 1              \\
\textbf{107} & 3000         & 0            & 1            & 0            & 1              & \textbf{144} & 1000         & 0            & 1            & 0            & 1              \\
\textbf{108} & 1000         & 1            & 0            & 0            & 1              & \textbf{144} & 2000         & 0            & 0            & 2            & 2              \\
\textbf{108} & 3000         & 0            & 0            & 1            & 1              & \textbf{145} & 1000         & 0            & 1            & 0            & 1              \\
\textbf{109} & 2000         & 0            & 0            & 1            & 1              & \textbf{146} & 1000         & 1            & 0            & 1            & 2              \\
\textbf{109} & 3000         & 0            & 1            & 0            & 1              & \textbf{146} & 3000         & 0            & 0            & 1            & 1              \\
\textbf{110} & 1000         & 1            & 0            & 1            & 2              & \textbf{147} & 1000         & 0            & 1            & 0            & 1              \\
\textbf{111} & 1000         & 1            & 0            & 0            & 1              & \textbf{148} & 3000         & 0            & 0            & 1            & 1              \\
\textbf{112} & 1000         & 0            & 1            & 0            & 1              & \textbf{149} & 1000         & 1            & 0            & 0            & 1
\end{tabular}
\end{table}

\begin{table}[ht]
\begin{tabular}{cccccccccccc}
\textbf{ID}  & \textbf{MIL} & \textbf{FM1} & \textbf{FM2} & \textbf{FM3} & \textbf{TOTAL} & \textbf{ID}  & \textbf{MIL} & \textbf{FM1} & \textbf{FM2} & \textbf{FM3} & \textbf{TOTAL} \\
\textbf{150} & 1000         & 1            & 0            & 0            & 1              & \textbf{164} & 3000         & 0            & 0            & 1            & 1              \\
\textbf{151} & 1000         & 0            & 0            & 1            & 1              & \textbf{165} & 1000         & 0            & 2            & 2            & 4              \\
\textbf{152} & 1000         & 1            & 0            & 0            & 1              & \textbf{165} & 2000         & 0            & 1            & 1            & 2              \\
\textbf{153} & 1000         & 0            & 1            & 0            & 1              & \textbf{165} & 3000         & 0            & 1            & 1            & 2              \\
\textbf{154} & 3000         & 1            & 0            & 0            & 1              & \textbf{166} & 1000         & 1            & 0            & 1            & 2              \\
\textbf{155} & 1000         & 0            & 1            & 0            & 1              & \textbf{167} & 1000         & 0            & 1            & 0            & 1              \\
\textbf{156} & 3000         & 0            & 1            & 0            & 1              & \textbf{167} & 3000         & 0            & 1            & 0            & 1              \\
\textbf{157} & 2000         & 0            & 0            & 1            & 1              & \textbf{168} & 1000         & 0            & 1            & 0            & 1              \\
\textbf{158} & 3000         & 0            & 0            & 1            & 1              & \textbf{169} & 1000         & 1            & 0            & 0            & 1              \\
\textbf{159} & 1000         & 0            & 0            & 1            & 1              & \textbf{169} & 2000         & 0            & 0            & 4            & 4              \\
\textbf{160} & 3000         & 0            & 0            & 1            & 1              & \textbf{169} & 3000         & 0            & 1            & 0            & 1              \\
\textbf{161} & 1000         & 0            & 1            & 2            & 3              & \textbf{170} & 1000         & 0            & 1            & 0            & 1              \\
\textbf{161} & 2000         & 0            & 1            & 2            & 3              & \textbf{170} & 2000         & 0            & 0            & 1            & 1              \\
\textbf{161} & 3000         & 1            & 0            & 2            & 3              & \textbf{171} & 1000         & 0            & 0            & 1            & 1              \\
\textbf{162} & 2000         & 1            & 0            & 0            & 1              & \textbf{172} & 2000         & 0            & 0            & 1            & 1              \\
\textbf{163} & 1000         & 0            & 1            & 0            & 1              & \textbf{}    &              &              &              &              &
\end{tabular}
\end{table}
\end{document}